\documentclass[aps,prl,twocolumn,showpacs,superscriptaddress]{revtex4}

\usepackage{graphicx}
\usepackage{color}
\usepackage{amsfonts}
\begin{document}
\title{Topological surface states on Bi$_{1-x}$Sb$_x$: Dependence on surface orientation, surface termination and stability}

\author{Xie-Gang Zhu}
\affiliation{Department of Physics and Astronomy, Interdisciplinary Nanoscience Center, Aarhus University,
8000 Aarhus C, Denmark}
\affiliation{Science and Technology on Surface Physics and Chemistry Laboratory, Mianyang 621700, People's Republic of China}
\author{Philip~Hofmann}
\affiliation{Department of Physics and Astronomy, Interdisciplinary Nanoscience Center, Aarhus University,
8000 Aarhus C, Denmark}
\email[]{philip@phys.au.dk}

\date{\today}

\begin{abstract}
Topological insulators support metallic surface states whose existence is protected by the bulk band structure. It has been predicted early that the topology of the surface state Fermi contour should depend on several factors, such as the surface orientation and termination and this raises the question to what degree a given surface state is protected by the bulk electronic structure upon structural changes. Using tight-binding calculations, we explore this question for the prototypical topological insulator Bi$_{1-x}$Sb$_x$, studying different terminations of the (111) and (110) surfaces. We also consider the implications of the topological protection for the (110) surfaces for the semimetals Bi and Sb. 
\end{abstract}
\maketitle

The defining feature of a topological insulator (TI) is the existence of metallic surface states that is brought about by certain topological properties of the insulator's bulk band structure. These topologically guaranteed surface states have a number of remarkable properties, for example their spin texture, and the scattering and localisation behaviour derived from it \cite{Zhang:2008,Moore:2010,Hasan:2010}. The surface states are often said to be ``protected'' by the bulk band structure and it is interesting to explore what this protection actually means. One aspect is the protection against certain scattering events, such as 180$^{\circ}$ backscattering. Another is the protection against destruction by disorder \cite{Hatch:2011,Schubert:2012} or structural rearrangements and this will be the focus of the present work.

It has been pointed out early on by Teo, Fu and Kane (TFK) that the detailed surface state topology of a TI depends on the surface orientation, and the general Fermi surface topology was discussed for some surface orientations and terminations of Bi$_{1-x}$Sb$_x$ \cite{Teo:2008}. According to TFK, the topology of the surface Fermi surface for a TI with inversion symmetry can be worked out by calculating the surface  fermion parity $\pi(\Lambda_a)$ at the surface time-reversal invariant momenta (TRIMs).  $\pi(\Lambda_a)$ can be evaluated 
from the  parity invariants $\delta$ at the bulk TRIMs $\Gamma_{i}$ by
\begin{equation}
\pi(\Lambda_a)=(-1)^{n_b}\delta(\Gamma_{i})\delta(\Gamma_{j}),
\label{eqn:equ1}
\end{equation}
where $n_b$ is the number of occupied, spin-degenerate bulk bands \cite{Teo:2008}. The $\pi(\Lambda_a)$ values then provide the Fermi contour topology. For two surface TRIMs, $\Lambda_a$ and $\Lambda_b$, the product $\pi(\Lambda_a)\pi(\Lambda_b)$ can be used to predict the number of surface state Fermi level crossings on a path connecting these surface TRIMs. For $\pi(\Lambda_a)\pi(\Lambda_b)=1(-1)$ an even (odd) number of crossings can be expected. Moreover, $\pi(\Lambda_a)$ as such can be used to determine the number of Fermi contours enclosing a given surface TRIM. For $\pi(\Lambda_a)=-1(1)$, the surface TRIM $\Lambda_a$ is enclosed by an odd (even) number of closed Fermi contours. 

\begin{figure}
 \includegraphics[width=0.4 \textwidth]{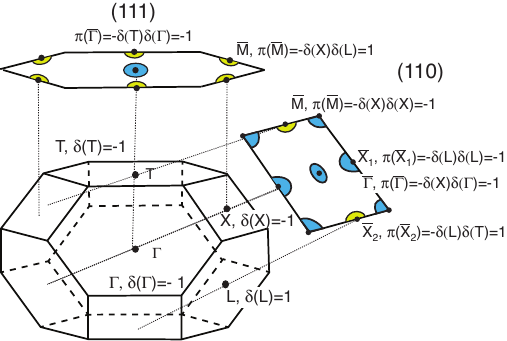}\\
\caption{(Color online) Topological predictions for the surface electronic structure of Bi$_{1-x}$Sb$_{x}$(111) and (110) ($ x \gtrsim 0.09$). The bulk Brillouin zone and the projected surface Brillouin zones are shown for both surface orientations. The black dots mark the positions of  time-reversal invariant momenta (TRIMs). The calculation of the surface fermion parity values $\pi$ for the surface TRIMs according to (\ref{eqn:equ1}) is illustrated. The blue areas denote an odd number of closed Fermi contours around a TRIM as derived from the surface Fermion parity in (\ref{eqn:equ1}). The yellow areas would be expected for the first atomic layer removed from either surface.  }
  \label{fig:1}
\end{figure}

The application of this is illustrated in Fig. \ref{fig:1} for the (111) and (110) surfaces of  Bi$_{1-x}$Sb$_x$. Essentially the same figure can also be used for the  Bi$_2$Se$_3$ class of materials by changing the name of the bulk TRIMs such that the TRIMs  $\Gamma$, L, X and T for  Bi$_{1-x}$Sb$_x$  are replaced by  $\Gamma$, L, F and Z for Bi$_2$Se$_3$, respectively. Evidently, the surface fermion parity values depend on the orientation of the surface because different bulk TRIMs are projected out onto different surface TRIMs, depending on the surface orientation. 

The application of (\ref{eqn:equ1}) is very simple for the Bi$_2$Se$_3$ class of materials, the most studied topological insulators. In this case, the bulk parity inversion happens only at the $\Gamma$ point. Together with the fact that the number of bulk bands $n_b$ is even (14), this leads to $\pi(\bar{\Gamma})=-1$ for \emph{every possible surface orientation}, since the bulk $\Gamma$ point is always projected out onto the surface $\bar{\Gamma}$ point. All the other surface TRIMs, on the other hand, have  $\pi(\Lambda_a)=1$. Consequently, the $\bar{\Gamma}$ point should be encircled by a closed Fermi contour for every possible surface orientation of the Bi$_2$Se$_3$ class and this has been confirmed for the (111) \cite{Xia:2009,Chen:2009} and (221) \cite{Xu:2013} orientations so far. Differences in the electronic structure of different surface orientations are important when considering the transition from one crystal face to another on e.g. a Bi$_2$Se$_3$ nano wire. One would expect a smooth transition since the electronic structure of the all the faces is essentially the same.

The situation is much more interesting for Bi$_{1-x}$Sb$_x$ because there the bulk parity invariants are negative for every TRIM except L. For the (111) surface, the L point projects out to the $\bar{M}$ point and one might therefore expect a negative surface fermion partity there but the application of (\ref{eqn:equ1}) actually gives the same result as for Bi$_2$Se$_3$: $\pi(\bar{\Gamma})=-1$ and $\pi(\bar{M})=1$. The reason is the odd number of bulk bands ($n_b=5)$ in Bi$_{1-x}$Sb$_x$ which changes the sign of all the $\pi$ values. The expected surface state topology for Bi$_{1-x}$Sb$_x$(111) is thus indicated by the blue areas in Fig. \ref{fig:1}. It is the same as for Bi$_2$Se$_3$(111) and this has also been confirmed experimentally \cite{Hsieh:2008}. The similarity of Bi$_2$Se$_3$(111) and Bi$_{1-x}$Sb$_x$(111), however, is a mere coincidence. In contrast to Bi$_2$Se$_3$, changing the surface orientation of Bi$_{1-x}$Sb$_x$ does change the electronic structure qualitatively and for  Bi$_{1-x}$Sb$_x$(110) three surface TRIMs were found to be encircled by closed Fermi contours \cite{Zhu:2013b}. This raises interesting questions about the transition from one crystal face to another for a Bi$_{1-x}$Sb$_x$ nano wire. It is also relevant for the design of more complex electronic structures on TI surfaces, such as the possibility to achieve a quasi one-dimensional electronic structure on a two-dimensional TI \cite{Wells:2009}. 

Apart from the surface orientation, the microscopic termination could play an interesting role for the surface electronic structure. This does not emerge directly from  (\ref{eqn:equ1}) but is rather hidden in the assumptions this equation is based on: When applying  (\ref{eqn:equ1}), it is assumed the that surface termination plane contains the same bulk inversion centre that has also been used to calculate the parity invariants $\delta(\Gamma_{i})$. If this is not so and the surface is terminated with different bulk inversion centres, the signs for the surface fermion parities may have to be changed. TFK have shown that a surface termination where such a change is necessary is the (111)' termination, a surface where the top half of the outermost bilayer of the A7 structure is removed \cite{Teo:2008}. Instead of the single blue Fermi contour encircling $\bar{\Gamma}$, one should expect to find the three yellow contours encircling the $\bar{M}$ points. Detailed changes of the surface electronic structure as a function of surface termination have been calculated for TI thin films \cite{Soriano:2012}, the topological crystalline insulator SnTe \cite{Liu:2013c,Safaei:2013} and discussed for SmB$_6$ \cite{Zhu:2013c}. The absence of such changes upon the simultaneous presence of different terminations has also been found for PbBi$_4$Te$_7$ \cite{Eremeev:2012b}. 

In this paper, we shall explore the dependence of the surface band structure on surface orientation and termination for the model TI  Bi$_{1-x}$Sb$_x$ using tight-binding based Green's function calculations for a semi-infinite solid. This approach has the advantage of avoiding any coupling between the two surfaces in a slab-type calculation. This can be particularly important for the non-(111) surfaces for Bi-like materials where the surface states can penetrate very deeply \cite{Hofmann:2005a}.  We use the tight-binding parameters of Liu and Allen \cite{Liu:1995}, interpolated for the alloy as in Ref. \cite{Teo:2008}. The calculations for the alloy are carried out for $x=0.14$. The bulk band structure projections are obtained by a direct projection of tight-binding bands onto the surface of interest.
For the surface state band structure calculation, a $sp^3$ Slater-Koster  tight-binding model is used \cite{Slater:1954}, which includes third-neighbour hopping parameters and spin-orbit coupling. Based on this, the surface states dispersion is calculated using a transfer-matrix technique and a Green's function approach \cite{Lee:1981b,Mele:1978}. The Green's function of the semi-infinite surface is generated from the transfer matrix which is calculated self-consistently. Its imaginary part represents the surface state dispersion as observed in ARPES experiments. Calculations for surfaces terminated by half a bilayer are implemented by setting the hopping matrix elements between the top layer and second layer in the first bilayer to zero and changing the on-site energy of the first layer atoms to a high value, in order to move their spurious spectral contribution out of the energy range of interest. Note that the calculations reported here are not derived from first principles but are instead based on the bulk tight-binding parameters. No structural optimisation can be performed, the resulting surface state dispersion can be significantly different from that obtained by a first principles calculation and no overall charge neutrality is guaranteed. However, for the present purpose of examining the topological properties of the surface states, the approach is still very useful. 

\begin{figure}
 \includegraphics[width=0.5 \textwidth]{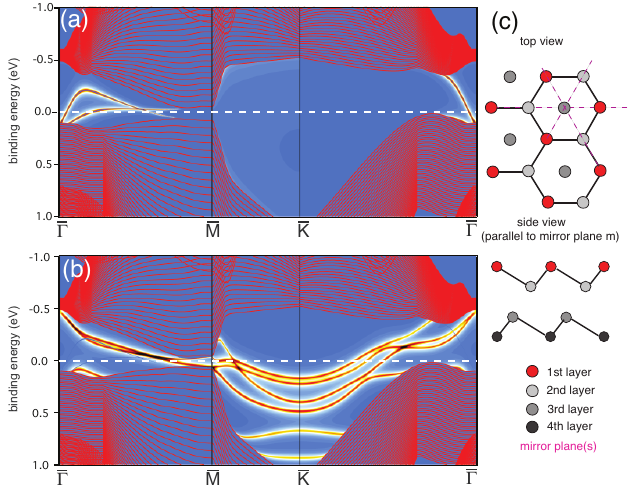}\\
\caption{(Color online) (a) Calculated surface electronic structure for Bi$_{0.86}$Sb$_{0.14}$(111). The red lines are the projected bulk band structure and the image is the imaginary part of the Green's function. (b) The same for Bi$_{0.86}$Sb$_{0.14}$(111)'. (c) Surface structure for the two surfaces. For the (111)' surface the (red) top layer atoms are missing. The black lines are the nearest neighbour bonds. }
  \label{fig:2}
\end{figure}

The result of such a calculation is shown for Bi$_{1-x}$Sb$_x$(111) in Fig. \ref{fig:2}(a). As expected,  it completely agrees with the similar calculation by TFK in Ref. \cite{Teo:2008} (see Figure 3 of that paper). It is easy to see that the result fulfils the topological predictions shown in Fig. \ref{fig:1} for the (111) surface. There is an odd number of Fermi level crossings between the surface TRIMs $\bar{\Gamma}$ and $\bar{M}$ and, moreover, $\bar{\Gamma}$, the TRIM for which $\pi=-1$, is encircled by an odd number of Fermi contours. The topologically required parity change between $\bar{\Gamma}$ and $\bar{M}$ is achieved by the two bands starting in the valence band at  $\bar{\Gamma}$, as one of these bands ends in the conduction band at $\bar{M}$ whereas the other stays in the valence band. Along $\bar{\Gamma}-\bar{M}$, a crossing between these bands is found that is not topologically required. This crossing is enforced by a subtle error in the Liu and Allen \cite{Liu:1995} tight-binding parameters that gives an incorrect mirror Chern number for this direction \cite{Teo:2008}.

Fig. \ref{fig:2}(b) shows the corresponding calculation for Bi$_{1-x}$Sb$_x$(111)', the surface with half the outermost bilayer (the red atoms in Fig. \ref{fig:2}(c)) removed. In contrast to the (111) surface, the cleavage plane of  (111)' does not contain the $c_0$ bulk inversion centre but the $c_1$ centre (for a definition of the inversion centres see Ref. \cite{Teo:2008}) and this is predicted to cause a sign change in all the surface fermion parities, such that the yellow Fermi contour shape in Fig. \ref{fig:1} is expected. The dispersion in Fig. \ref{fig:2}(b) shows a drastic change of the surface electronic structure upon this switch of termination. Most importantly, the surface states that disperse from $\bar{\Gamma}$ to $\bar{M}$ now start out in the conduction band, not in the valence band. The topologically required parity switching still holds as one of the bands ends in the valence band and one in the conduction band and also the spurious crossing in between the bands is observed. However, this change in termination  illustrates that the topological protection of the surface states only holds for the overall expectation of metallic states in between certain TRIMs. It does not give protection for any specific state if there is a possibility of surface structural changes.

\begin{figure}
 \includegraphics[width=0.5 \textwidth]{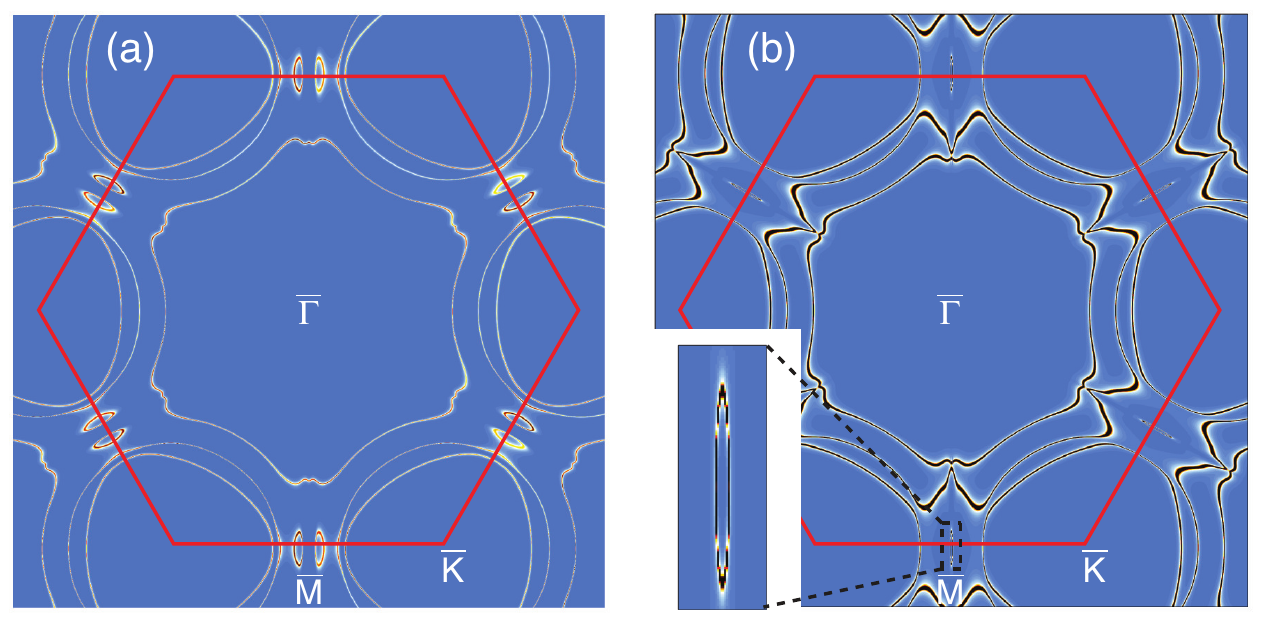}\\
\caption{(Color online) (a) Fermi contour for Bi$_{0.86}$Sb$_{0.14}$(111)' together with the surface Brillouin zone (or imaginary part of the Green's function taken at the Fermi energy). (b) The same but with the Fermi level assumed to be 27~meV higher. This shift results in a closed pocket around $\bar{M}$ that is magnified in the inset.}
  \label{fig:3}
\end{figure}

The topological predictions for Bi$_{1-x}$Sb$_x$(111)' do not only imply an odd number of crossings between $\bar{\Gamma}$ and $\bar{M}$, they also imply that $\bar{M}$ is encircled by an odd number of Fermi contours whereas $\bar{\Gamma}$ is encircled by none or an even number. This is difficult to infer from Fig. \ref{fig:2}(b) because of the many new states around $\bar{K}$, a high symmetry point that is not a TRIM. Fig. \ref{fig:3}(a) therefore shows the calculated Fermi contour for Bi$_{1-x}$Sb$_x$(111)'. This Fermi contour is not consistent with the topological prediction because it shows one closed Fermi contour around $\bar{\Gamma}$ and none around $\bar{M}$, even though the number of Fermi level crossings between these TRIMs is still odd, as also seen in the dispersion of Fig. \ref{fig:2}(b). Fig. \ref{fig:3}(a) also shows a number of closed contours around $\bar{K}$ and along the $\bar{K}-\bar{M}$ line but these do not have any topological significance. 

The problem can be resolved by moving the Fermi energy upward by 27~meV, a value that is still within the gap of the TI at this composition. This results in the Fermi contour in Fig. \ref{fig:3}(b). Now one finds two closed contours around $\bar{\Gamma}$ and one around $\bar{M}$, albeit a very small one. The failure of the simple calculation to reproduce the  expected Fermi contour topology is not surprizing because the tight-binding scheme employed here does not guarantee charge neutrality, as pointed out in connection with the calculational methods. 

\begin{figure}
 \includegraphics[width=0.5 \textwidth]{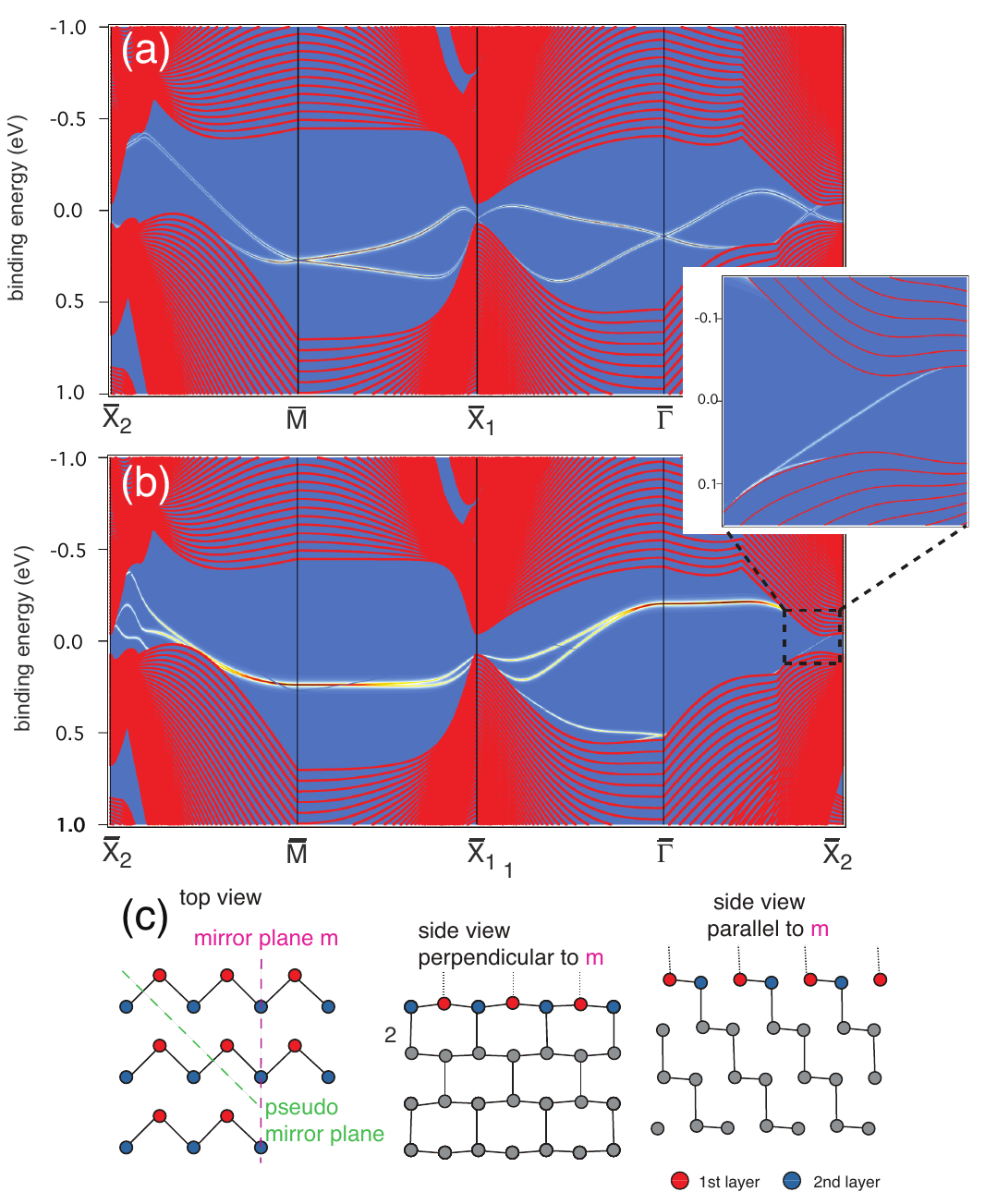}\\
\caption{(Color online) (a) Calculated surface electronic structure for Bi$_{0.86}$Sb$_{0.14}$(110). The red lines are the projected bulk band structure and the image is the imaginary part of the Green's function. (b) The same for Bi$_{0.86}$Sb$_{0.14}$(110)' with the electronic structure near the $\bar{X}_2$ point magnified in the inset. (c) Structure of the (110) surface. For the (110)' surface the (red) first layer atoms are missing. The green line (pseudo mirror line) would be a mirror line if the crystal structure was simple cubic instead of rhombohedral A7.  The black lines are the nearest neighbour bonds. }
  \label{fig:4}
\end{figure}

\begin{figure}
 \includegraphics[width=0.5 \textwidth]{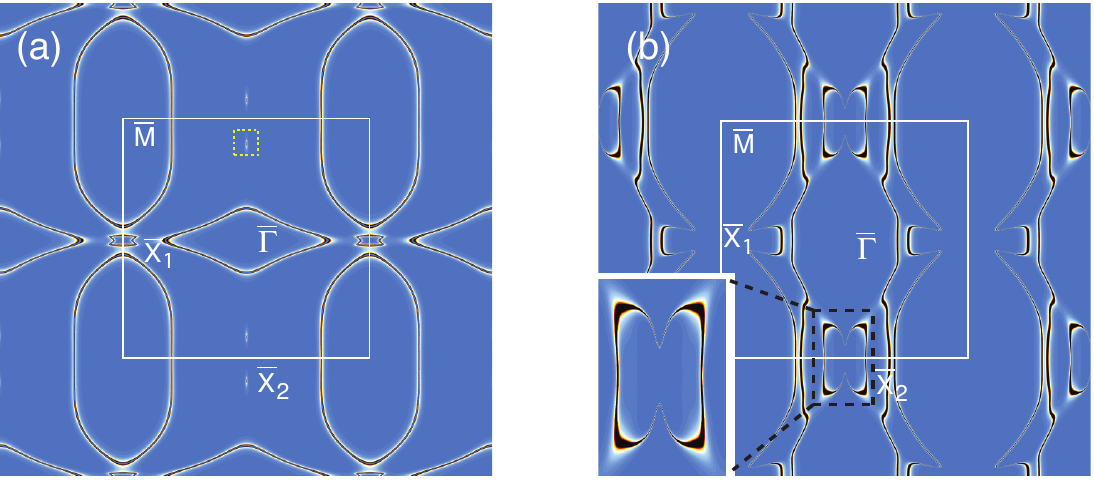}\\
\caption{(Color online) (a) Fermi contour for Bi$_{0.86}$Sb$_{0.14}$(110) together with the surface Brillouin zone (or imaginary part of the Green's function taken at the Fermi energy). The yellow square contains a very small electron pocket. (b) The same for  Bi$_{0.86}$Sb$_{0.14}$(110)' with the hole pocket around $\bar{X}_2$ magnified in the inset. }
  \label{fig:5}
\end{figure}

We now turn to another surface termination of Bi$_{1-x}$Sb$_x$ that has been studied experimentally, the (110) surface. The calculated band structure is shown in Fig. \ref{fig:4}(a). As already discussed in Ref. \cite{Zhu:2013b} where this calculation is taken from, the qualitative agreement of this calculation with the topological predictions shown in Fig. \ref{fig:1} is excellent but the quantitative agreement is notably poorer than for the (111) surface. In particular, the closed contours around $\bar{\Gamma}$ and $\bar{M}$ are found to be electron pockets in the Green's function calculation but hole pockets in the experiment and also in first principles calculations for the very similar Bi(110) \cite{Koroteev:2004,Pascual:2004}. As for the (111) surface, a spurious crossing of the bands in the mirror direction of the surface ($\bar{\Gamma}-\bar{X}_2$) is observed, caused by the above-mentioned reason of the incorrect mirror Chern number. The same good qualitative agreement with the topological predictions and the experimental result is found for the Fermi contour in Fig. \ref{fig:5}(a) where the spurious crossing gives rise to a very small and topologically irrelevant electron pocket along $\bar{\Gamma}-\bar{X}_2$.

It is also possible to find a different surface termination for the (110) direction which we call (110)'. A side view on the geometric structure of the (110) surface (see Fig. \ref{fig:4}(c)) shows that the atoms in the first two layers have almost the same height. As for the (111) surface, removing the first layer atoms causes the surface to be terminated by different inversion centres than the one used to calculate the bulk parity values $\delta$ ($c_1,c_2,c_{23},c_{13}$ instead of $c_0,c_3,c_{12},c_{123}$) and this changes the sign of all the surface fermion parities $\pi$. This should lead to the yellow Fermi contour in Fig. \ref{fig:1}, i.e. an odd number of closed contours around $\bar{X}_2$ only. Note that the removal of atoms on this surface is expected to cause the opposite transition  from the (111) surface:  For the (111) surface, a removal of one layer changes the electronic structure from one TRIM enclosed by Fermi contours to three but for (110) the change is the other way around. 

The surface state dispersion for Bi$_{1-x}$Sb$_x$(110)' is shown in Fig. \ref{fig:4}(b) and the corresponding surface Fermi contour in Fig. \ref{fig:5}(b). As for the (111) surface termination change, we also observe a change of the electronic structure here but it appears less drastic. At first glance, and apart from some more pronounced changes around $\bar{X}_2$, the dispersion of the states appears shifted and the spin-splitting is reduced. This is particularly clear around $\bar{\Gamma}$ and $\bar{M}$ where there is hardly any splitting of the spin-degenerate states in the vicinity of the surface TRIMs. Moreover, the dispersion of the states along the $\bar{\Gamma}-\bar{X}_2$ and $\bar{M}-\bar{X}_1$ is very small, suggesting that these states are largely localised along the mirror plane of the structure (shown in Fig. \ref{fig:4}(c)). 

The Fermi contour topology of Bi$_{1-x}$Sb$_x$(110)' agrees with the prediction of merely one closed contour around $\bar{X}_2$ (see Fig. \ref{fig:1}). The state generating this feature is a hole pocket, stemming from a band dispersing down from the conduction band to the valence band in the vicinity of $\bar{X}_2$. The three closed Fermi contours around the other TRIMs on  Bi$_{1-x}$Sb$_x$(110) are removed: The upper spin-split band around $\bar{X}_1$ is moved down in energy, removing the closed contour around this TRIM, and the pockets around $\bar{M}$ and $\bar{\Gamma}$ are removed by an additional change of surface state exchanges between valence band and condition band in the $\bar{M}-\bar{X}_2$ and $\bar{\Gamma}-\bar{X}_2$ directions, respectively.

The Fermi contour of Bi$_{1-x}$Sb$_x$(110)' in Fig. \ref{fig:5}(b) does also reflect the confinement of the states in the direction of the mirror plane (i.e. $\bar{\Gamma}-\bar{X}_2$). The states that form the pockets around $\bar{\Gamma}$ and $\bar{M}$ for the (110) surface are still there but they now form quasi one-dimensional sheets of Fermi contour that do not surround any surface TRIM. There appear to be two factors causing this confinement. First, the loss of top (red) atoms in the surface unit cell in Fig. \ref{fig:4}(c) evidently must reduce the dispersion along the mirror line because a state dispersing in this direction would involve hopping matrix elements over the red atom. Moreover, the dispersion along the mirror line is anyway expected to be smaller than perpendicular to the mirror line because it involves hopping of electrons in between the bilayers making up the A7 structure. These bilayers are easily identified in the side view parallel to the mirror plane of Fig. \ref{fig:4}(c). Finally, the reduced spin-orbit splitting for (110)' also contributes to a smaller band width. Overall, the apparent reduction of dimensionality that is observed for the transition between the (110) and (110)' surfaces could be indicative for the mechanism that leads to completely one-dimensional states on the vicinal surfaces of Bi \cite{Wells:2009}.

\begin{figure}
 \includegraphics[width=0.5 \textwidth]{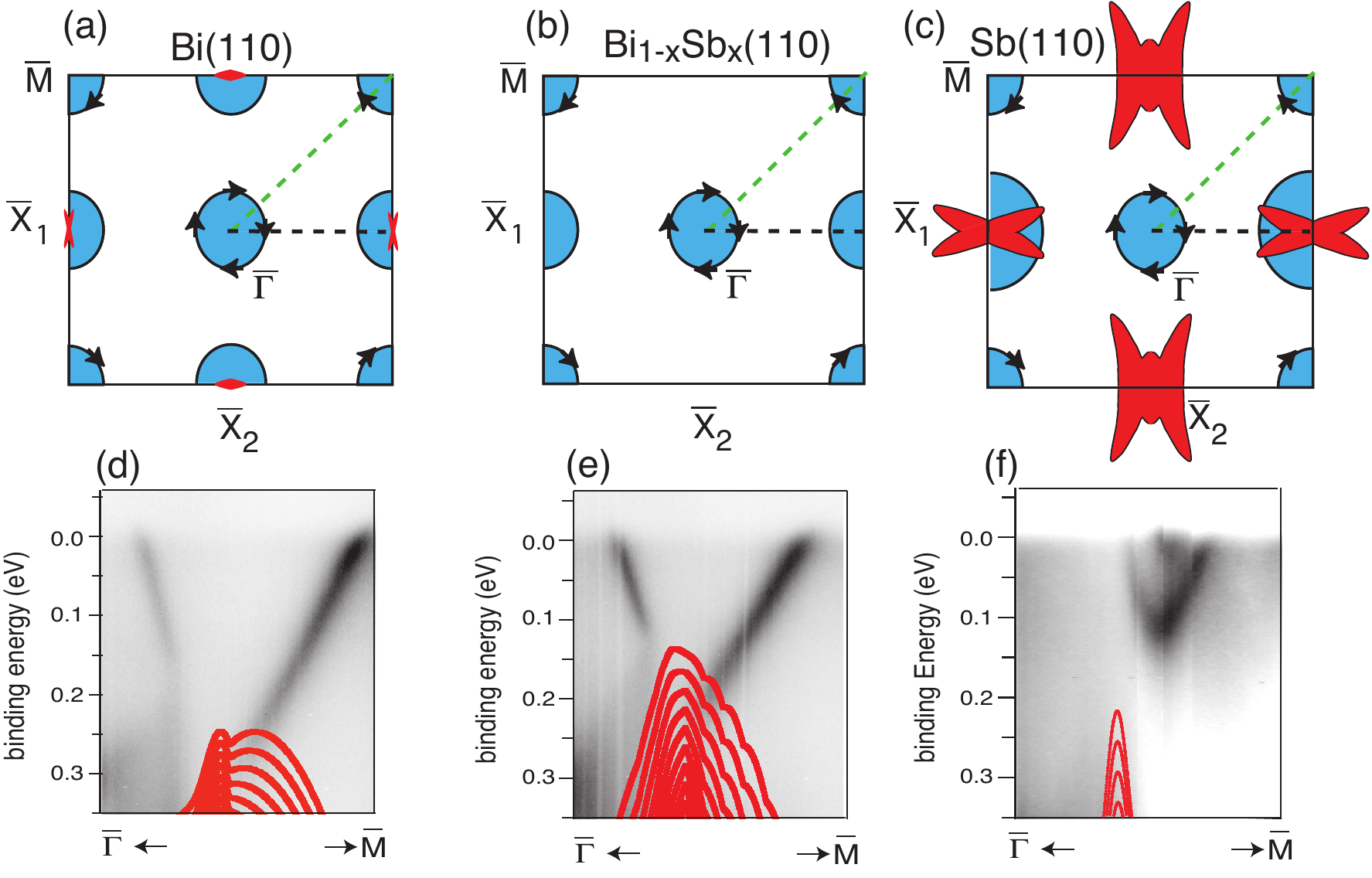}\\
\caption{(a)-(c) Prediction for the surface state topology for the (110) surfaces of Bi, Bi$_{1-x}$Sb$_x$ and Sb. The blue areas denote an odd number of closed Fermi contours around a TRIM. The red areas are the projected bulk Fermi surface. The arrows on the Fermi contours around $\bar{\Gamma}$ and $\bar{M}$ denote the calculated spin directions. (d)-(f) ARPES measurements of the electronic structure of these surfaces along the $\bar{\Gamma}-\bar{M}$ line (dashed line in (a)-(c)). Dark corresponds to high photoemission intensity. The black lines are the projection of the bulk bands onto the (110) surface. Data for (a) and (b) taken from Ref. \cite{Zhu:2013b}, for (c) from Ref. \cite{Bianchi:2012}. }
  \label{fig:6}
\end{figure}

We now address a slightly different issue in connection with the (110) and ask how strongly the topology protects the surface states even if the underlying surface is not a topological insulator or not even has a topologically non-trivial band structure. Fig. \ref{fig:6} (a)-(c) show the expected surface state topology for the (110) surfaces of Bi, Bi$_{1-x}$Sb$_x$ and Sb. These have been determined ignoring the fact that Bi and Sb are actually semimetals, not insulators. Note that, as a result of the bulk band structure topology, the predicted surface state topology is the same for Bi$_{1-x}$Sb$_x$ and Sb but different from Bi. Both Bi$_{1-x}$Sb$_x$ and Sb have $\delta=-1$ for every bulk TRIM except L, whereas for Bi $\delta$ is also $-1$ for the bulk L point. Therefore, one obtains a topologically non-trivial Fermi surface with three closed contours around surface TRIMs for Bi$_{1-x}$Sb$_x$ and Sb whereas the topologically trivial band structure of Bi leads to four closed Fermi contours, an even number.

To what degree can the topological predictions be used for the surfaces of the semimetals Bi and Sb? Consider a specific direction in $k$-space such as $\bar{\Gamma}-\bar{X}_1$ (black dashed line in Fig. \ref{fig:6} (a)-(c)). For all three surfaces, both $\bar{\Gamma}$ and $\bar{X}_1$ are encircled by a closed Fermi contour or, more precisely, by an odd number of such contours, such that there is an even number of Fermi level crossings between $\bar{\Gamma}$  and $\bar{X}_1$. However, even though $\pi(\bar{\Gamma})\pi(\bar{X}_1)=1$, the situation along this line is not topologically trivial and the states cannot be removed. Indeed, since both $\pi$ values are -1, both surface TRIMs have to be enclosed by a Fermi contour. On the other hand, both Bi and Sb are semimetals and any path from $\bar{\Gamma}$ to $\bar{X}_1$ has to go through a piece of projected Fermi surface. This strongly limits the validity of the topological predictions because any required parity or symmetry change between valence band and conduction band could happen within the projected Fermi surface without having to involve surface states. A similar situation arises for most high-symmetry directions on Bi and Sb, especially also for the $\bar{\Gamma}-\bar{M}$ direction on Bi(111) for which the strict application of topological predictions has been discussed \cite{Ohtsubo:2013}.

The situation is different along  $\bar{\Gamma}-\bar{M}$ for the (110) surfaces (green dashed line in Fig. \ref{fig:6} (a)-(c)). Again, two closed Fermi contours are predicted, one around each of the two surface TRIMs but now the two TRIMs can be connected by a line that does not pass through any segment of projected bulk Fermi surface. The topological prediction should thus be a firm one. The experimental results in Fig. \ref{fig:6}(d)-(f) appear to confirm this: Experimentally, the states around both surface TRIMs are hole pockets and one always finds two crossings due to these. For Sb(110), one finds two more crossings along $\bar{\Gamma}-\bar{M}$ due to a small electron pocket along this line.  For Bi and Bi$_{1-x}$Sb$_x$, the two hole pocket-derived bands do not join each other but disperse into the bulk bands. For Sb(110), they do appear to join without reaching the bulk band projection but one has to keep in mind the limited accuracy of the tight-binding calculation used for the band structure projection. 

When inspecting the  spin texture along the two hole pockets (arrows on the Fermi contour in Fig. \ref{fig:6}(a)-(c) as calculated for Bi(110) \cite{Pascual:2004,Strozecka:2011} and Sb(110) \cite{Bianchi:2012,Strozecka:2012} and here assumed to be the same for the alloy), it appears obvious that the two contours cannot be joint because the spin rotates in opposite directions along the contours. However, this picture is too simple because the spin is not conserved in strongly spin-orbit coupled systems such as these here. In fact, detailed calculations of the spin polarisation along the $\bar{\Gamma}-\bar{M}$ line reveal that the spin polarisation is present near the Fermi level crossings but lost about mid-way between $\bar{\Gamma}$ and $\bar{M}$ \cite{Strozecka:2011,Strozecka:2012}, such that it would be possible to connect the states, as apparently found for Sb(110). 

Apart from the absence of a projected bulk Fermi contour along $\bar{\Gamma}-\bar{M}$, the stabilisation of the topological states on Bi(110) and Sb(110) is aided by two factors. Especially for Bi(110), the spin-orbit interaction is very strong and this leads to a situation where the states have to be degenerate at the surface TRIMs but split strongly away from these points \cite{Hofmann:2006}. Moreover, the geometric structure of Bi and Sb leads to a quasi mirror-protection of the states, similar to the situation on a topological crystalline insulator \cite{Fu:2011}. The A7 rhombohedral structure of Bi is very close to being a simple cubic structure. This can easily be seen in the structure of the (110) surface shown in Fig. \ref{fig:4}(c) as a very small lattice distortions would be sufficient to turn the structure into simple cubic. The rhombohedral (110) surface would then be a pseudocubic (100) surface  \cite{Jona:1967,Hofmann:2006}. In this case, the green dashed line in Fig. \ref{fig:4}(c) would be a mirror plane of the structure and the states in the corresponding $\bar{\Gamma}-\bar{M}$ direction could be classified by their mirror symmetry along this plane. This would make it impossible to join the hole pocket around the $\bar{\Gamma}$  with that around the $\bar{M}$ point because the mirror symmetry of the states would need to be strictly conserved, unlike the spin.

A few important conclusions can be drawn from this work. As expected, we find metallic surface states on all investigated surface orientations and terminations of the topological insulator Bi$_{1-x}$Sb$_x$, and the Fermi surface topology of these states is also consistent with the detailed predictions by TFK \cite{Teo:2008}.  However, a change of surface termination by removing one layer of atoms, to obtain (111)' from (111) and (110)' from (110), leads to drastic changes not only in the surface state Fermi contour topology but also in the surface electronic structure. For the (111) orientation, for instance, it changes the electronic structure near the $\bar{\Gamma}$ point such that the topological states originate from the conduction band instead of the valence band. This shows that while the metallic surfaces are protected, as expected, specific surface states are not protected and depend on the local surface geometric structure. It should be emphasised that a cut-bilayer termination may be so unstable that it might not be possible to ever realise it. This is very likely to be true for the (111) surface and probably also for (110). For the (100) orientation of Bi, however, metastable terminations between bilayers have been observed  \cite{Sun:2010}.

A second conclusion concerns the design freedom that arises from the possibility to influence the surface electronic structure via the crystal termination. The removal of the first layer of atoms to obtain (111)' from (111) and (110)' from (110) involves a change in the number of surface TRIMs that are enclosed by closed Fermi contours from one to three and vice versa. This transition is particularly illustrative for the (110) orientation because several of the bands from the (110) termination are still present, they only change character from being closed around surface TRIMs to being quasi one-dimensional, not enclosing any TRIMs. This illustrates how the topologically required changes of surface Fermi contours between different facets of a crystal could happen and how quasi one-dimensional topological states on vicinal surfaces could be created, as in the case of Bi(114) \cite{Wells:2009}. 

Finally, we have seen how surface states on non-topological semimetals like Bi can also benefit from the topological protection, especially Fermi contour segments between two surface TRIMS that can be connected along a path not crossing any parts of the projected bulk Fermi surface. Such a protection is limited by the possibility to change the surface termination such that all surface fermion parities would change sign, something that can potentially destroy \emph{all} the topological surface states on a trivial insulator or a semimetal.

We gratefully acknowledge stimulating discussions with Shuichi Murakami as well as financial support by the VILLUM foundation.


\end{document}